\documentclass[prl,aps,twocolumn,superscriptaddress,showpacs,floatfix]{revtex4}
\usepackage{graphicx}
\usepackage{amsmath}

\begin{document}

\title{Edge effects in the two-dimensional spin-$\frac{1}{2}$ Heisenberg antiferromagnet}

\author{Kaj H. H\"oglund} 
\affiliation{Department of Physics, {\AA}bo Akademi University, 
Porthansgatan 3, FI-20500, Turku, Finland}

\author{Anders W. Sandvik}
\affiliation{Department of Physics, Boston University, 590 Commonwealth Avenue,
Boston, Massachusetts 02215}

\date{\today}


\begin{abstract}
We use quantum Monte Carlo simulations to study effects of free edges in the two-dimensional spin-$\frac{1}{2}$ Heisenberg antiferromagnet. We find that the magnetic 
response of an edge is smaller than the bulk susceptibility. This counter-intuitive quantum effect can be traced to enhanced antiferromagnetic nearest-neighbor 
spin correlations, i.e., tendency to local singlet formation, at and close to the edge. These correlations form a comb-like pattern, which can be reproduced with a simple 
variational valence-bond state. We also study rough edges, and find that these instead significantly enhance the susceptibility, due to local sublattice imbalance
impeding singlet formation.
\end{abstract}

\maketitle

Among the intricacies of strongly-correlated quantum systems, the roles of various defects, such as impurities and boundaries, are intriguing. Defects inevitably effect 
experiments, to an extent often not precisely known. On the other hand, they can also serve as useful experimental probes of correlated quantum states 
\cite{mahajan,takigawa1,bobroff,julien}. Theoretically, impurity effects in quantum antiferromagnets have been studied extensively in one \cite{eggert1,sirker} 
and two dimensions \cite{bulut,nagaosa,sachdev,hoglund,anfuso,metlitski1}. Impurities can effectively cut isolated spin chains into finite segments with 
free ends, which leads to particularly strong deviations from bulk magnetic properties in one dimension \cite{eggert1}. A good example is the quasi-one-dimensional 
antiferromagnet Sr$_2$CuO$_3$, for which the NMR line exhibits a broad background \cite{takigawa1} which is well accounted for by the local magnetic susceptibility (Knight-shift) 
distribution of open-end Heisenberg chains \cite{takigawa1,eggert1}. In two dimensions, free edges can be expected to have less dramatic consequences, because of the 
typically small ratio of boundary to bulk, and not much attention have been paid to them. However, with the increasing focus on nano-scale materials, 
the boundary physics should become accessible (or unavoidable, depending on the perspective) also in two-dimensional antiferromagnets. It is therefore important to establish 
what edge effects to expect based on prototypical model systems such as the Heisenberg hamiltonian. In this Letter we take some steps in this direction.

We use the approximation-free stochastic series expansion (SSE) quantum Monte Carlo method \cite{sse} to study the antiferromagnetic ($J>0$) Heisenberg hamiltonian,
\begin{equation}
H = J\sum _{\left <i,j\right >}\mathbf{S}_{i}\cdot \mathbf{S}_{j},
\label{eq:hamiltonian}
\end{equation}
where $\mathbf{S}_i$ are the usual $S=\frac{1}{2}$ spin operators and $\langle i,j\rangle$ denotes nearest-neighbors on a square $L\times L$ lattice. We consider systems
with completely open boundaries as well as semi-open ones, which are periodic in one direction and open in the other direction. The absence of corners and the translational 
symmetry along the open edges in the semi-open systems allow easier access to an infinite edge. On the other hand, it is also interesting to study corner effects in the 
fully open systems. Experimentally, it is likely that typical samples would have some roughness, leading to effects not captured by the 
smooth edges of the $L \times L$ lattices. We therefore also study systems with irregular edges, constructed according to a scheme described further below. We are interested 
in the magnetic response of the edges, and will also investigate how this is related to changes in the spin-spin correlations relative to those in fully periodic systems.

\begin{figure}
\includegraphics[width=6.5cm,clip]{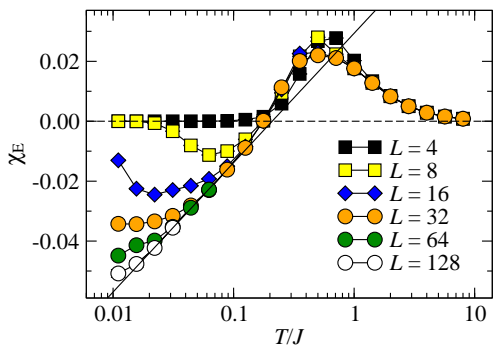}
\null\vskip-1mm
\caption{(Color online) Edge susceptibility of semi-open systems. Error bars are smaller than the symbols. The line is a log-lin fit to the size-converged data.}
\label{fig:fig1}
\vskip-3mm
\end{figure}

In analogy with the impurity susceptibility previously 
considered for systems with isolated vacancies or added spins \cite{sachdev,hoglund} we define an {\it edge susceptibility},
\begin{equation}
\chi_{\rm E} = \frac{\chi_a - \chi_0 }{aL},
\label{eq:edge}
\end{equation}
where $\chi_a$ is the total magnetic susceptibility for a system with $a$ free edges; $a=0$ for periodic systems, $a=2$ for semi-open boundaries, and
$a=4$ for fully open systems. In all cases there are $N=L^2$ spins and the total susceptibilities are given by
\begin{equation}
\chi_a = \frac{1}{T} \left \langle M_z^2\right \rangle,~~~M_z=\sum_{i=1}^{N} S^z_i.
\end{equation}
The normalization with $aL$ in (\ref{eq:edge}) reflects the natural assumption that the difference in response should scale with the total length of the edges
of the open or semi-open systems. One would intuitively anticipate $\chi _{\rm E}>0$, as the edge spins should be free to fluctuate more than those in the bulk.  
Surprisingly, this actually does not hold when $T\ll J$ and $L \to \infty$. Fig.~\ref{fig:fig1} shows the temperature dependence of $\chi_{\rm E}$ for semi-open 
systems. The scaling of $\chi_2 - \chi_0$ with the edge length is confirmed as the $\chi_{\rm E}$ curves for different $L$ coincide for increasing $L$. For $T\gg J$, 
the spins contribute independently $(4T)^{-1}$ to the susceptibilities and, consequently, $\chi_{\rm E}$ vanishes. In the limit $T/J\to 0$, 
we must have $\chi_{\rm E} \to 0$ for any (even) $L$ because the ground state is a singlet regardless of the boundary condition (i.e., $M_z=0$), which 
is seen explicitly for $L=4$, and $8$. Focusing on the $L \to \infty$ converged data, decreasing $T$ initially leads to an increasing $\chi_{\rm E}$, in line with the 
expectation of $\chi_2 > \chi_0$ due to enhanced fluctuations of the edge spins. However, a maximum is reached at $T/J\approx 0.5$, below which $\chi_{\rm E}$ 
decreases and becomes negative. The temperature dependence below $T/J \approx 0.1$ is consistent with a logarithmic divergence; $\chi_{\rm E} \propto -\ln(J/T)$.
This behavior can be contrasted with the single-impurity susceptibility, which, for both vacancies and added spins, is always positive when $T \to 0$, diverging 
as a Curie form with a log correction \cite{sachdev,hoglund}. In the case of the edge problem considered here, where the defect is non-magnetic (the number of spins 
is the same in the periodic and open systems), there is no {\it a priori} reason to expect a divergent $\chi_{\rm E}$. We have also run simulations for 
the classical Heisenberg antiferromagnet with semi-open boundaries and in that case find $\chi _{\rm E}$ to converge to a positive constant as $T\to 0$. Hence, the 
negative divergent edge susceptibility has to be attributed to quantum effects beyond those included in the ``renormalized classical'' description \cite{chn} of the 
Heisenberg model. 

\begin{figure}
\includegraphics[width=6.4cm,clip]{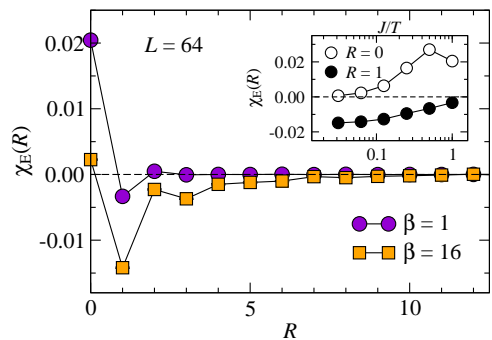}
\null\vskip-1mm
\caption{(Color online) Position-dependent edge susceptibility at distance $R$ from the free edge of a $64\times 64$ semi-open lattice at two 
different temperatures. The temperature convergence for $R=0,1$ is shown in the inset.}
\label{fig:fig2}
\vskip-3mm
\end{figure}

When normalized by the total number of spins $L^2$, instead of the edge length $L$, the divergent edge susceptibility would be a negligible correction to the bulk 
susceptibility, which is constant as $T \to 0$ \cite{chn}. However, in principle a local susceptibility can be accessed in NMR experiments through the Knight shift
\cite{knight,takigawa1}, and, provided that sufficient sensitivity can be achieved and the edges are smooth enough (both of which may clearly pose challenges), the 
divergence should be detectable. We thus also study the spatially resolved susceptibility, defined for a site $i$ as $\chi_{a}(i) = \beta \langle S_{i}^{z}M_z\rangle$. 
For a periodic system, $a=0$, there is no dependence on the location $i$, whereas in the semi-open system $\chi_{2}(i)$ depends only on the distance $R$ of $i$ from 
the edge; in either case $\sum_i \chi_a(i)=\chi_a$. We define the position dependent edge susceptibility,
\begin{equation}
\chi_{\rm E}(R)  = \chi_{2}(R) - \chi_{0}/L^{2},
\end{equation}
which obeys $\sum_R\chi_{\rm E}(R)=\chi_{\rm E}$. Fig.~\ref{fig:fig2} shows this quantity at two different temperatures at which the results are size-converged for 
$L=64$. Here it is seen that a large contribution to the negative edge susceptibility comes from the second line of spins from the edge, $R=1$. The response of the 
edge line, $R=0$, is always larger than the bulk value, however, and $\chi_{\rm E}(0)$ seems to vanish, or become very small, as $T \to 0$. Beyond $R\approx 10$, 
$\chi_{\rm E}(R)$ becomes very small, indistinguishable from zero, also for the lower temperature. The available data suggest that $\chi_{\rm E}(R)$ should become 
negative for any $R$ at sufficiently low $T$ (and $L \to \infty$). Considering the log-divergent $\chi_{\rm E}$, we should have $\chi_{\rm E}(R) \propto R^{-1}$ when 
$T \to 0$.

We will next show that the negative edge susceptibility is related to enhanced local spin correlations close to the edge. The spin correlators 
$\langle {\bf S}_i \cdot {\bf S}_j\rangle$ for nearest-neighbor sites $i,j$ (bonds) form a pattern of weak and strong bonds. These correlations
provide a measure of the amplitude of spins $i,j$ forming a singlet. In addition to calculating the correlations with the SSE method, we have also used 
a variational state in the valence-bond basis, from which some additional insights are gained. 

A valance-bond basis state for $N$ spins is a product of $N/2$ 
singlets $(a,b)=(\uparrow_{a}\downarrow_{b}-\downarrow_{a}\uparrow_{b})/\sqrt{2}$, where $a$ and $b$ are sites on different sublattices of the bipartite square 
lattice. Any singlet state $|\Psi\rangle$ can be expanded in this over-complete basis;
\begin{equation}
|\Psi\rangle = \sum_{v}\psi(v)|(a^v_1,b^v_1)\cdots(a^v_{N/2},b^v_{N/2})\rangle, 
\end{equation}
where $v \in \{1,\ldots \frac{N}{2}!\}$ labels the different bond configurations. In the amplitude-product state of Liang {\it et al.} \cite{liang}, 
the wave function coefficients are products of real amplitudes $h(a,b)$;
\begin{equation}
\psi(v) = \prod_{i=1}^{N/2}h(a^v_i,b^v_i).
\end{equation}
\noindent
\begin{figure}
\includegraphics[width=5.5cm,clip]{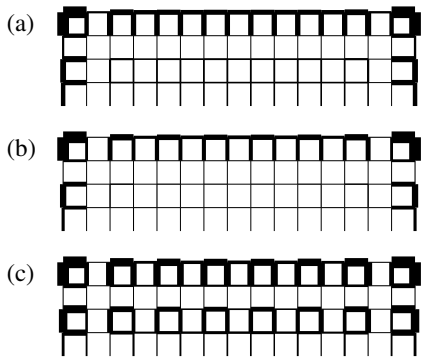}
\null\vskip-1mm
\caption{Bond patterns at an edge of an $L=16$ system obtained with (a) an optimized amplitude-product state (almost exact), (b) a state with amplitudes 
$h(r)={\rm e}^{-r}$, and (c) classical dimers. The line widths correspond to $-\langle {\bf S}_i \cdot {\bf S}_j\rangle$ in the range $[0.315,0.451]$ and 
$[0.316,0.466]$ in (a) and (b), respectively, and average dimer occupation $\in [0.167,0.500]$ in (c).}
\label{fig:fig3}
\vskip-3mm
\end{figure}
\noindent
For a periodic system the amplitudes depend only on the bond lengths (the $x$ and $y$ separations of the two sites), $h(a^v_i,b^v_i)=h(x^v_i,y^v_i)$, but in an 
open system they depend on both site coordinates (up to reflection and rotation symmetries). It is known that the periodic Heisenberg model can be very well 
described by this simple state \cite{liang,lou}. The amplitudes decay as $h(r) \sim r^{-3}$, where $r$ is the bond length \cite{lou}. To study boundary (including 
corner) effects, we have optimized all amplitudes for a $16\times 16$ fully open lattice, using the optimization method discussed in \cite{lou}. Also in this 
case the amplitude-product state  provides a very good description of the system, with the energy deviating by less than $0.1$\% from the approximation-free result 
obtained using the SSE method at very low $T$. The bond pattern is also almost identical in the variational and SSE calculations; the result is shown in 
Fig.~\ref{fig:fig3}(a). The correlations are the strongest at the corners, and at the edge they form a comb-like pattern. This comb-structure is repeated on alternating 
columns away from the boundary, with a rapidly decaying amplitude, as shown in Fig.~\ref{fig:fig4} based on SSE calculations. The strongest bonds are enhanced by more 
than $10\%$ compared to the bulk. This enhancement is associated with higher amplitudes for local singlets within the combs, which is clearly consistent with the 
reduced edge susceptibility. The correlation modulations are seen in Fig.~\ref{fig:fig4} to decay more rapidly with $R$ than the edge susceptibility in Fig.~\ref{fig:fig2}. 
On the other hand, the nearest-neighbor correlations represent only the dominant contribution to local singlet formation, and thus the two properties are not easily related 
to each other quantitatively.

\begin{figure}
\includegraphics[width=6.5cm,clip]{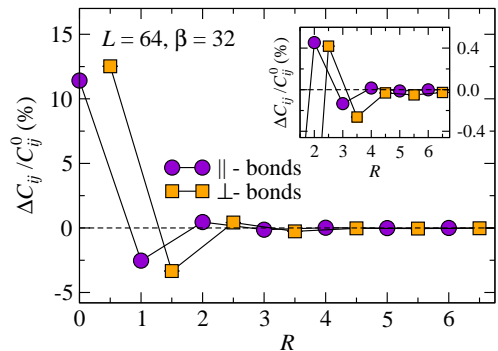}
\null\vskip-1mm
\caption{(Color online) Deviation $\Delta C_{ij}=C_{ij}-C_{ij}^0$ of the nearest-neighbor correlation $C_{ij}=|\langle {\bf S}_i \cdot {\bf S}_j\rangle|$ 
from the bulk value $C_{ij}^0$ versus the distance $R$ from an edge of an $L=64$ semi-open system at $T=J/32$. Integer and half-integer $R$ correspond to 
bonds $ij$ parallel and perpendicular to the edge, respectively. The inset shows magnified $R\ge 2$ data.}
\label{fig:fig4}
\vskip-3mm
\end{figure}

In order to elucidate the origin of the comb-like bond pattern, as well as the plaquette structure at the corners, we have calculated the correlations also in a non-optimal 
state where the bond-amplitudes are very short-ranged, $h(r)={\rm e}^{-r}$, which corresponds to a spin liquid \cite{liang}. In spite of this state being very different from 
the actual N\'eel ground state of the Heisenberg model, a very similar bond pattern forms at the corners and edges of the open lattice, as shown in Fig.~\ref{fig:fig3}(b). 
This indicates that the edge pattern is rather insensitive to the long-range correlations of the state---it is essentially governed by the hard-core nature of short valence 
bonds. To further illustrate this point, we show in Fig.~\ref{fig:fig3}(c) the bond-occupation pattern of the classical dimer model (with only short bonds, averaging over bond 
configurations using Monte Carlo sampling). Here there is more of a tendency to plaquette formation at the edge, which, however, changes into a uniform comb-structure 
away from the corners of larger lattices. These results show that the gross features of the boundary correlation pattern of the Heisenberg model is dominated by the entropy 
of short valence bonds, and thus we argue that this is at the heart also of the reduced edge susceptibility. 

The site dependent susceptibility $\chi _{a}(i)$ can be used to derive experimental consequences, such as the NMR line-shape \cite{takigawa1,anfuso}. However, it is 
unclear whether the edge effects we have discussed so far could be observed experimentally. Samples consisting of extremely small fragments are most likely required 
to distinguish any edge features from the NMR bulk signal, and the fragments appearing in powders hardly have long smooth edges; more likely they have irregular shapes. 
In principle one could carry out simulations for a suitable ensemble of clusters. However, not knowing the actual structure of clusters that could be expected 
in experiments, we here consider a simple model for roughness added to the $L\times L$ open systems discussed above, with the aim of studying the robustness of the 
smooth-edge effects.

\begin{figure}
\includegraphics[width=6.25cm,clip]{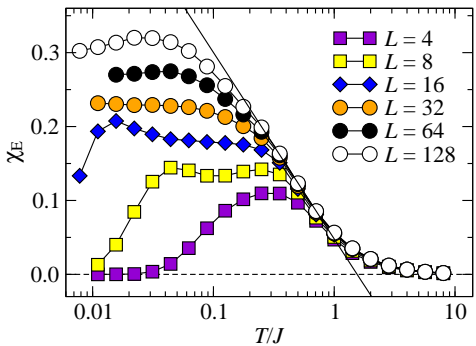}
\null\vskip-1mm
\caption{Edge susceptibility for systems with rough edges. The line is a fit to the size-converged intermediate-$T$ data.}
\label{fig:fig5}
\vskip-3mm
\end{figure}

Our roughness model amounts to traversing the $4(L-1)$ boundary sites of an $L\times L$ lattice and removing each spin with probability $p$ or coupling a new spin 
to it with probability $p$ (and doing nothing with probability $1-2p$). In order to have the same number of spins in the periodic $L\times L$ systems and these 
rough-boundary system, we only study samples with the same number of added and removed spins, and, furthermore, we only consider clusters with equal numbers of 
spins on both sublattices (so that the total ground-state spin remains $0$). The impurity susceptibility $\chi_{\rm E}$, Eq.~(\ref{eq:edge}), then still vanishes for finite 
$L$ both in the limits $T \to 0$ and $T \to \infty$. Figure~\ref{fig:fig5} shows results averaged over several hundred random boundaries for each $L$. Here $p=1/3$, 
corresponding to maximum roughness. We again observe a non-trivial logarithmic divergence, but, in contrast with the smooth edge, $\chi_{\rm E}$ is always positive. 
The prefactor is about $5$ times larger than in the smooth-edge case. Thus we conclude that the roughness has completely changed the nature of the edge effect. 
For less rough boundaries (smaller $p$) the prefactor of the log divergence is reduced, and for some very small $p$ we expect to recover the negative factor 
pertaining for the smooth boundaries. We have not yet carried out systematic studies of this, however. 

Within the picture of the reduced susceptibility of the smooth edge arising from local singlets, the different response of the rough edges can be understood
as an impeded singlet (short valence bond) formation due to local sublattice imbalance. This has been previously examined in diluted systems (randomly placed 
vacancies) \cite{wang}. In that case, as the percolation point is approached local sublattce imbalance leads to localized magnetic moments, which interact and 
form low-lying states at an energy scale below the normally lowest-lying ``quantum rotor'' states of the Heisenberg antiferromagnet on finite clusters. Here, for the
rough-edge problem, we have only studied the static magnetic response, but it would clearly be interesting to study also other aspects of the rough boundaries, 
e.g., their excitations.

In summary, we have found that smooth open edges in the two-dimensional $S=\frac{1}{2}$ Heisenberg model have a smaller magnetic response than the bulk, contrary to the 
naively expected enhancement of fluctuations due to the smaller number of neighbors of the edge spins. We have explained this surprising effect in terms of local singlet 
formation at the edges, which in turn can be regarded as a consequence of entropy maximization in a valence-bond description of the system. In sharp contrast to smooth edges, 
rough boundaries lead to an enhanced magnetic susceptibility. We have argued that this is due to local sublattice imbalance (``dangling spins''), which impedes local 
singlet (short valance bond) formation. For both smooth and rough boundaries, the edge susceptibility of an infinite system diverges logarithmically as $T \to 0$. These 
studies also demonstrate that edges should have profound effects on the magnetic response of nano-scale clusters, and that details of the boundary texture are important. 

The smooth-edge effects that we have pointed out here have very recently been examined using field-theoretical methods by Metlitski and Sachdev \cite{metlitski2}. The negative 
edge susceptibility originates from low-lying spin waves. The prefactor of the log-divergence, the slope in Fig.~\ref{fig:fig1}, is in reasonable agreement with the prediction 
of \cite{metlitski2}. The comb-structure in the edge correlations was argued to be a short-distance phenomenon, as we have also shown here, beyond the standard O(3) 
continuum filed theory description. It can be understood in terms of proximity to a phase transition into a valence-bond-solid state. 

We would like to thank Max Metlitski, Subir Sachdev, and Masashi Takigawa for stimulating discussions. 
The work of AWS was supported by the NSF under grant No.~DMR-0513930.

\null

\end{document}